\documentclass[twocolumn,showpacs,amsmath,amssymb,superscriptaddress]{revtex4}

\newtheorem{theorem}{Theorem}
\newtheorem{corollary}{Corollary}

\begin{document}

\title{Popescu-Rohrlich correlations as a unit of nonlocality}

\author{Jonathan Barrett}
\email{jbarrett@perimeterinstitute.ca}
\affiliation{Perimeter Institute for Theoretical Physics, 31 Caroline Street N, Waterloo, Ontario N2L 2Y5, Canada}

\author{Stefano Pironio}
\email{spironio@caltech.edu}
\affiliation{Institute for Quantum Information, California Institute of Technology, Pasadena, CA 91125, USA}

\date{June 19, 2005}

\begin{abstract}
A set of nonlocal correlations that have come to be known as a PR box suggest themselves as a natural unit of nonlocality, much as a singlet is a natural unit of entanglement. We present two results relevant to this idea. One is that a wide class of multipartite correlations can be simulated using local operations on PR boxes only. We show this with an explicit scheme, which has the interesting feature that the number of PR boxes required is related to the computational resources necessary to represent a function defining the multipartite box. The second result is that there are quantum multipartite correlations, arising from measurements on a cluster state, that cannot be simulated with $n$ PR boxes, for any $n$.
\end{abstract}

\maketitle
By performing measurements on an entangled quantum system, two separate observers can obtain correlations that are nonlocal, in the sense that the joint probabilities $P(a_1a_2|x_1x_2)$ for the observers to get the outcomes $a_1$ and $a_2$ given the measurements $x_1$ and $x_2$ cannot be written in the product form $P(a_1a_2|x_1x_2)=\sum_j p_j P_j(a_1|x_1)P_j(a_2|x_2)$ with $p_j\geq 0$ and $\sum_j p_j=1$  \cite{bel87}. The nonlocal character of the correlations implies that two parties who wish to simulate the experiment with classical resources only cannot do so without communication. Nonlocal correlations, although they cannot be used to signal from one observer to the other, can be exploited in various information processing tasks, such as in communication complexity \cite{bra03}, or for the distribution of a secret key between two parties \cite{bhk04}.

Nonlocality can thus be viewed as an information-theoretic resource and investigated as such \cite{blm05}. Forty years after Bell's seminal paper, however, we still lack a proper theoretical framework --- analogous, e.g., to the framework that has been developed for the study of entanglement --- that would allow us to answer unambiguously questions such as, can two given sets of nonlocal correlations be considered equivalent resources, or, what is a good measure of nonlocality. In particular, we have not yet identified what would constitute a unit of nonlocality, in the same way that the singlet state constitutes the unit of entanglement.

To progress on these issues, drawing analogies with entanglement is a natural way to proceed. But note a conceptual difference between entanglement and nonlocality: while entanglement is intimately related to the tensorial structure of quantum mechanics, nonlocality, on the contrary, can be defined without reference to quantum theory. In particular, it is possible to write down sets of nonsignaling correlations that are more nonlocal than allowed by quantum mechanics \cite{kt85}. Why is quantum mechanics not more nonlocal \cite{pr94}? What are the implications of the quantum restrictions, and what are the principles at the origin of these restrictions \cite{vda05}? A quantitative approach to nonlocality, as required in an information-theoretic perspective, may help us answer these questions.

In this spirit, it is useful to consider nonlocal correlations in the abstract, and not necessarily as arising from a set of measurements on a quantum state. Suppose that two observers have access to a black box. When an observer $i$ introduces an input $x_i$, the box produces an output $a_i$. The box is characterized by the joint probability $P(a_1a_2|x_1x_2)$ of obtaining the output pair $(a_1,a_2)$ given the input pair $(x_1,x_2)$. Compatibility with special relativity requires that these joint probabilities satisfy the nosignaling conditions
\begin{equation}\label{nosig}
\sum_{a_2}P(a_1a_2|x_1x_2)=\sum_{a_2}P(a_1a_2|x_1x'_2)\equiv P(a_1|x_1)
\end{equation}
for all $a_1,x_1,x_2,x'_2$, as well as a similar set of conditions obtained by summing over the first observer's outputs. This ensures that one observer cannot signal to the other via his choice of input in the box. Apart from these constraints, the joint distribution can be arbitrary, and in particular nonlocal. The definition of nonlocal boxes generalizes to more parties in a straightforward way.

Some comparisons with entangled quantum states are as follows. Nonlocal boxes and entangled states both represent undirected resources that can be shared between two or more parties. In both cases, the set of allowed states is convex. Extremal elements of this convex set can be thought of as pure states, whereas others are mixed states. There is a notion of monogamy of nonlocality analogous to the monogamy of entanglement \cite{blm05}.

Entanglement theory is based on the notion that the entanglement contained in different quantum states can be compared by interconverting them through local operations and classical communication (LOCC), where a basic premise is that LOCC cannot on average increase the entanglement. In the case of bipartite pure states, reversible interconversion is possible, at least asymptotically, and this leads to a unique measure of entanglement. In the case of multipartite states, or bipartite mixed states, reversibility is not in general possible, and this makes the situation more complicated. But still, we do have that any entangled state (bipartite or multipartite) can be obtained from sufficient copies of the singlet state. Similarly, it is possible to study interconversions between nonlocal boxes, i.e., the simulation of nonlocal boxes using other boxes as a resource. In this context, the parties are allowed unlimited access to shared randomness and have the ability to perform local operations, such as relabelling inputs and outputs, or using the output for one box as the input for another box. Communication, however, is not allowed, since it enables trivially the simulation of any nonlocal box. We can now ask the following question. Is there an elementary nonlocal box that allows the simulation of all other boxes, and which could thus be viewed as a unit of nonlocality?

A plausible candidate is the following box, which was described by Kalfhi and Tsirelson \cite{kt85}, and was independently introduced in a more physical context by Popescu and Rohrlich \cite{pr94}. It takes two inputs $x_1,x_2\in\{0,1\}$ and produces two outputs $a_1,a_2\in\{0,1\}$ according to the joint distribution
\begin{equation}\label{xy}
P(a_1a_2|x_1x_2) = \left\{\begin{array}{l@{\quad}l} 1/2:&
a_1+a_2={x_1x_2} \mod 2\\
0:& \mathrm{otherwise.}
\end{array} \right.
\end{equation}
The marginals thus satisfy $P(0|x_1)=P(1|x_1)=1/2$ and similarly for $x_2$. By Tsirelson's theorem \cite{tsi80}, these correlations cannot be obtained from measurements on any quantum state. Following the denomination used in other works, we will refer to this box as a Popescu-Rohrlich (PR) box. The PR box is maximally nonlocal for the class of two-input two-output boxes, the simplest class of nonlocal boxes \cite{blm05}, and is a natural primitive for communication complexity as it allows the solution of any problem with one bit of communication \cite{vda05}. In analogy, the singlet state is the maximally entangled state of two qubits, the simplest family of entangled states, and is a natural primitive for entanglement consuming information processing tasks, such as teleportation or dense coding.

To answer our previous question, it is important to determine if all nonlocal correlations can be obtained from PR boxes, or if they are any which cannot. It was shown in \cite{blm05}, that all two-input bipartite boxes can be simulated with PR boxes (at least in an approximate sense). It is also known that one PR box is sufficient to reproduce correlations arising from arbitrary von Neumann measurements on the singlet state \cite{cgm04}. Other examples of correlations that can be simulated with PR boxes are given in \cite{bm05}.

In this article, we investigate further the potential of the PR box. We first present a simple protocol that allows the simulation of a large class of two-output boxes. These boxes are natural generalizations of the PR box to more inputs and more parties. It follows from our construction that any $n$-partite communication complexity problem can be solved with $n-1$ bits of communication and a number of PR boxes related to the computational resources necessary to represent its objective function. A second consequence of our result is that any bipartite box with binary outputs can be simulated with PR boxes. We then consider a box that does not belong to the previous class. It arises from measurements of Pauli operators on a cluster state of five qubits. We demonstrate that the corresponding correlations cannot be simulated with PR boxes, or even with arbitrary bipartite boxes.

We now show how PR boxes can simulate a large class of multipartite correlations. These boxes are $n$-partite boxes with an arbitrary number of inputs for each party and with binary outputs. We denote the $n$-tuple of inputs as $\vec{x}=(x_1\ldots,x_n)$, where, without loss of generality, each input $x_i\in\{0,\ldots,2^m-1\}$ and can thus be represented by a $m$-bit string. The $n$-tuple of outputs is $\vec{a}=(a_1\ldots,a_n)$, where $a_i\in\{0,1\}$. We consider boxes characterized by the following joint distribution
\begin{equation}\label{ncorr}
P(\vec{a}|\vec{x}) = \left\{ \begin{array}{l@{\quad}l} 1/2^{n-1}:&
\sum_i a_i=f(\vec{x})\mod 2\\
0:& \mathrm{otherwise,}
\end{array} \right.
\end{equation}
where $f(\vec{x})$ is a Boolean function of the inputs. Note that the outputs for any subset of $n-1$ parties are completely random. The only nontrivial correlations thus involve the full set of $n$ parties. Boxes of the form (\ref{ncorr}) are the most general two-output boxes with this property. We will refer to them as full-correlation boxes.

\begin{theorem}
Any full-correlation box can be simulated with PR boxes.
\end{theorem}

\emph{Proof.}
Our proof uses the fact that a Boolean function can be represented as a Boolean circuit, and that the NAND gate, whose action on two input bits $q$ and $r$ is $\textrm{NAND}(q,r)=qr+1$, constitutes a universal gate for Boolean circuits \cite{pap94}. We begin by supposing that the $n$ parties have already succeeded in simulating a full correlation $n$-partite box with outputs $\beta_i$ such that $\sum_i \beta_i=g_1(\vec{x})$, and a full correlation $n$-partite box with outputs $\gamma_i$ such that $\sum_i \gamma_i=g_2(\vec{x})$. We show how, by using these two boxes along with $n(n-1)$ PR boxes, they can simulate a single $n$-partite box with outputs $a_i$ such that $\sum_i a_i=\mathrm{NAND}(g_1(\vec{x}),g_2(\vec{x}))$. The simulation of a general full-correlation box, using PR boxes only, consists of iterations of this basic building block - one for each NAND gate in a circuit that evaluates $f(\vec{x})$.

Suppose then that each party possesses two bits $\beta_i$ and $\gamma_i$, such that the $n$-bit strings $\vec{\beta}$ and $\vec{\gamma}$ satisfy respectively $\sum_i \beta_i=g_1(\vec{x})$ and $\sum_i \gamma_i=g_2(\vec{x})$. Each pair of parties shares two PR boxes between them (making a total of $n(n-1)$).
Let $B_{ij}$, $i\neq j$, denote a PR box shared between party $i$ and $j$. (Our notation is such that the two PR boxes shared between parties $i$ and $j$ are $B_{ij}$ and $B_{ji}$.) In box $B_{ij}$ party $i$ inputs $\beta_i$ and gets an output $b_{ij}$, while party $j$ inputs $\gamma_j$ and gets an output $c_{ij}$. It thus follows that $b_{ij}+c_{ij}=\beta_i\gamma_j$. The final output of party $i$ is given by  $a_i=\sum_{j\neq i} \left(b_{ij}+c_{ji}\right)+\beta_i\gamma_i+r_i$, where $r_i=1$ if $i=1$, $r_i=0$ otherwise. If we sum (modulo 2) the $n$ outputs we thus get
\begin{align}
\sum_ia_i&=\sum_i\sum_{j\neq i}\left(b_{ij}+c_{ji}\right)+\sum_i\beta_i\gamma_i+\sum_i r_i\nonumber\\
&=\sum_i\sum_{j\neq i}\left(b_{ij}+c_{ij}\right)+\sum_i\beta_i\gamma_i+\sum_i r_i\nonumber\\
&=\sum_i\sum_{j}\beta_i\gamma_j+1=\sum_i\beta_i\sum_j\gamma_j+1\nonumber\\
&=g_1(\vec{x})g_2(\vec{x})+1=\textrm{NAND}(g_1(\vec{x}),g_2(\vec{x}))\,.
\end{align}
Moreover, because the outputs of a PR box are locally completely random, the outputs for any subset of $n-1$ parties take each of the possible values in $\{0,1\}^{n-1}$ with equal probability. It follows that each possible value of $\vec{a}$ consistent with $\sum_i a_i=\textrm{NAND}(g_1(\vec{x}),g_2(\vec{x}))$ occurs with probability $1/2^{n-1}$ as required in (\ref{ncorr}).

Consider again a circuit comprised of NAND gates that evaluates $f(\vec{x})$, such that the inputs to the final NAND gate in the circuit are $g_1(\vec{x})$ and $g_2(\vec{x})$. If we assume that we can already simulate the boxes characterized by $g_1$ and $g_2$, then we can simulate the $f(\vec{x})$ box. But $g_1(\vec{x})$ is itself the output of a NAND gate with inputs $h_1(\vec{x})$ and $h_2(\vec{x})$, so we can simulate the $g_1$ box with the same construction. We keep working backwards until we reach the point where inputs to NAND gates are simply the input bits themselves. But the corresponding boxes are local and can be simulated without PR boxes. \hfill$\Box$

\begin{corollary}
Any $n$-partite communication complexity problem can be solved with $n-1$ bits of communication and at most $kn(n-1)$ PR boxes, where $k$ is the size of the smallest circuit comprised of NAND gates that computes $f(\vec{x})$.
\end{corollary}

In communication complexity, $n$ parties are each given an input $x_i$ and must compute a function $f(\vec{x})$ of their joint inputs while communicating as little as possible. This problem can easily be solved if the parties share a box of the form (\ref{ncorr}). It suffices that each party introduces his input into the box and communicates his output to the first party, who can recover the value of $f(\vec{x})$ by summing all the outputs. Corollary 1 then simply follows from Theorem 1.

\begin{corollary}
Any two-output bipartite box can be simulated with PR boxes.
\end{corollary}

We sketch the proof.

\emph{Proof:}
For a fixed number of inputs and outputs, the set of nonlocal bipartite boxes is a convex polytope \cite{blm05}. To simulate any two-output bipartite box, it is thus sufficient to simulate every box which is a vertex of the corresponding polytope, since the others can be obtained as mixtures of vertices. Further, it is sufficient to focus on genuine two-output boxes, that is on boxes such that for every input $x_1$, $P(0|x_1)>0$ and $P(1|x_1)>0$, and similarly for every input $x_2$. Indeed, if a box satisfies $P(0|x_1)=0$ or $P(1|x_1)=0$ for some $x_1$, it is straightforward that we can simulate this box if we can simulate the box obtained from it by removing input $x_1$. Finally, it is then easy to show that every genuine two-output extremal box is of the form (\ref{ncorr}). (This can be done, for example, by adapting a proof of \cite{blm05}, where the polytope of two-input $d$-output boxes is characterized.)\hfill$\Box$

We have noted that the outputs for any subset of fewer than $n$ parties are completely random in boxes of the form (\ref{ncorr}) and that these are the most general boxes with this property. We now give an example of correlations that are not of that form and that cannot be simulated with PR boxes. The fact that there exist nontrivial correlations between subsets of the parties is crucial to prove this fact. The correlations arise from spin measurements on a one dimensional cluster state of five qubits in a ring \cite{br01}. Cluster states are remarkable in that they can act as a universal substratum for measurement based computation \cite{rb01}, as well as playing a role in quantum error correction \cite{sw01}. Their nonsimulability by PR boxes is thus another interesting property.
The correlations can be described as follows. Let inputs $x=0$ and $x=1$ correspond to spin measurements in the $\sigma_z$ and $\sigma_x$ bases, and let $a_i$ be the output of party $i$ for input $x_i=0$, and $a'_i$ for input $x_i=1$. Then we have
\begin{equation}\label{corr1}
a_1+a'_2+a_3=0\mod 2 \,,
\end{equation}
and cyclic permutations of the parties, together with
\begin{equation}\label{corr2}
a'_1+a'_2+a'_3+a'_4+a'_5=1\mod 2\,.
\end{equation}
These correlations were described in \cite{dp97} and constitute a GHZ-type paradox. (If values are assigned locally to $a_i$ and $a'_i$ for each party, then on summing the left hand sides together one obtains $0$; summing the right hand sides, on the other hand, gives $1$.)

\begin{theorem}
The correlations of Eqs.~(\ref{corr1}) and (\ref{corr2}) cannot be simulated exactly by parties who share random data and $n$ PR boxes, for any $n$.
\end{theorem}

In order to prove this result, it is useful to describe how a general protocol that aims to simulate these correlations with boxes will work. Let the value of the shared random data on a given round be $\lambda$, with probability $P(\lambda)$. Consider a particular party, Alice say, and denote her measurement $x$. Suppose that Alice shares $m$ boxes with other parties. Label these boxes $B_1,\ldots, B_m$.
Alice proceeds as follows.
\begin{enumerate}
\item She puts an input $y_{1}$ into box $B_{i_1}$, where $i_1=i_1(\lambda,x)$, and $y_{1}=y_{1}(\lambda,x)$. She obtains an output $\alpha_{1}$.
\item She puts an input $y_{2}$ into box $B_{i_2}$, where $i_2=i_2(\lambda,x,\alpha_{1})$, and $y_{2}=y_{2}(\lambda,x,\alpha_{1})$. She obtains an output $\alpha_{2}$.
\item She continues in this fashion until all $m$ boxes have been used. Her final output is a function $a=a(\lambda,x,\alpha_{1},\ldots,\alpha_{m})$.
\end{enumerate}
Finally, a different strategy along these lines may of course be defined for each party.

Note that the nosignaling conditions ensure that the correlations $P(a_2\ldots a_n|x_2\ldots x_n)$ restricted to the other parties do not depend on the specifics of Alice's protocol. In particular they do not depend on wether Alice uses her boxes or not.

\emph{Proof of Theorem 2:}
We begin by supposing that there is a protocol for simulating the above correlations exactly using $n$ PR boxes. Then we show that this implies that there is a protocol for simulating the correlations exactly using shared random data alone. This we know to be impossible.

A protocol for simulating the correlations must of course produce outputs correlated according to the six equations described by (\ref{corr1}) and (\ref{corr2}). Consider the case in which Alice's measurement is $x_1=0$. She follows some strategy as described above. Suppose that there is some set of values of $\lambda,\alpha_{1},\ldots,\alpha_{{m-1}}$, occurring with non-zero probability, such that $a_1(\lambda,{x_1=0},\alpha_{1},\ldots,\alpha_{{m-1}},\alpha_{m}=0)\neq a_1(\lambda,{x_1=0},\alpha_{1},\ldots,\alpha_{{m-1}},{\alpha_{m}=1})$. Suppose that the $m^\mathrm{th}$ PR box in this sequence is shared with party 5. Eq. (\ref{corr1}) implies that the outputs of parties 1, 2 and 3 should be correlated according to $a_1+a'_2+a_3=0$. Since this equation does not involve party 5, we may as well assume, by nosignaling, that party 5 does not use her half of the box. Then $\alpha_m$ is random and uncorrelated with the rest of the protocol. This means that, conditioned on the specified values of $\lambda,\alpha_{1},\ldots,\alpha_{m-1}$ occurring, Alice has a $1/2$ chance of outputting a value $a_1$ that is not correctly correlated with the outputs of parties 2 and 3. Furthermore, it does not matter whom the $m^\mathrm{th}$ box is shared with. For any of the other parties, there is one equation (\ref{corr1}) such that Alice's input is $x_1=0$ and the other party is not involved.

We can conclude from the above that $a_1(\lambda,{x_1=0},\alpha_{1},\ldots,\alpha_{{m-1}},{\alpha_{m}=0})= a_1(\lambda,{x_1=0},\alpha_{1},\linebreak[4]\ldots,\alpha_{{m-1}},{\alpha_{m}=1})$ for all values of $\lambda,\alpha_{1},\ldots,\alpha_{{m-1}}$. This means that in fact, Alice never needs to know the value of $\alpha_{m}$, and her strategy may as well terminate before putting an input into the $m^\mathrm{th}$ box. But now we can run an identical argument for the $(m-1)^\mathrm{st}$ box, concluding that Alice's strategy terminates after the $(m-2)^{\textrm{nd}}$ box, and so on. We conclude that if errors are not tolerated, Alice's output, in the event her measurement is $x_1=0$, must be fixed by $\lambda$ alone, and she does not use her boxes. We can then run this argument for each of the five parties.

Finally, consider Alice's strategy in the event her measurement is $x_1=1$. Then her output $a' _1$ should satisfy the constraint $a'_1+a_2+a_5=0$. We have already established that in this case, parties 2 and 5 do not use PR boxes, and output values that depend only on $\lambda$. Thus Alice's output must also be deterministic, and fixed by $\lambda$. The PR boxes in her possession simply output values that are random and uncorrelated with the rest of the protocol, so she cannot use them. This argument may now be run for each of the five parties.

We have established that none of the five parties in fact use the PR boxes in their possession, for any of the measurements $x_i=0,1$. Thus if they are producing exactly the required correlations, they are doing so using only shared random data. This we know to be impossible, which concludes the proof.\hfill$\Box$

It is straightforward to modify the proof we have just given to show that the correlations (\ref{corr1}) and (\ref{corr2}) cannot be reproduced with any bipartite box (including boxes with more inputs or outputs than a PR box).

\emph{Conclusion.} We have shown that PR boxes can be used to simulate a large class of correlations. These include all bipartite boxes with binary outputs, and therefore, for example, von Neumann measurements on non-maximally entangled states of two qubits (Cerf et al. have shown that a single PR box can simulate a maximally entangled state \cite{cgm04}). This encourages the idea that PR boxes should be considered as a proper unit of bipartite nonlocality. In a multipartite scenario, we have seen that PR boxes cannot simulate correlations arising from measurements on a five-qubit cluster state.
One could consider the approximate simulation of these correlations, where one demands that the error can be made arbitrary small. It seems reasonable to think that even in this case, PR boxes cannot simulate the cluster state correlations we described, in which case they cannot be considered as a unit of nonlocality in a multipartite scenario.

It thus appears that the structure of nonlocal correlations is rather different from the structure of entanglement. There are many open questions, for example, are there $n$-partite boxes such that any set of $(n-k)$-partite boxes are not sufficient for their simulation? Can we define a finite set of boxes that would be sufficient for the simulation of all $n$-partite boxes? One obvious reason for the difference between entanglement and nonlocal boxes may be the fact that we have not allowed classical communication in our protocols for manipulating boxes. An interesting extension of these ideas would be to introduce secrecy, and to see how boxes can be transformed in the presence of public, but not private communication.

\emph{Note added.} After completion of this work, N. Jones and Ll. Masanes informed us that they independently derived the result of Corollary 2 \cite{mj05}.

\acknowledgements
S. P. acknowledges support by the David and Alice Van Buuren fellowship of the Belgian American Educational Foundation and by the National Science Foundation under Grant No. EIA-0086038.

\end{document}